\documentclass[aps,superscriptaddress,amsmath]{revtex4}

\bibpunct{(}{)}{;}{a}{,}{,}

\usepackage{graphicx}

\begin{document}

\title{Dispersion properties of electrostatic oscillations in quantum plasmas}

\author{Bengt Eliasson}

\affiliation{Institut f\"ur Theoretische Physik IV, Ruhr-Universit\"at Bochum, D-44780 Bochum, Germany}

\affiliation{Department of Physics, Ume{\aa} University, SE--901 87 Ume{\aa}, Sweden}

\author{Padma K. Shukla}

\affiliation{Institut f\"ur Theoretische Physik IV, Ruhr-Universit\"at Bochum, D-44780 Bochum, Germany}

\begin{abstract}
We present a derivation of the dispersion relation for electrostatic oscillations (ESOs) in a zero temperature 
quantum plasma.  In the latter, degenerate electrons are governed by the Wigner equation, while non-degenerate 
ions follow the classical fluid equations. The Poisson equation determines the electrostatic wave potential. 
We consider parameters ranging from semiconductor plasmas to metallic plasmas and electron densities of 
compressed matter such as in laser-compression schemes and dense astrophysical objects. Due to the wave diffraction
caused by overlapping electron wave function due to the Heisenberg uncertainty principle in dense plasmas, 
we have possibility of Landau damping of the high-frequency electron plasma oscillations (EPOs) at large enough 
wavenumbers. The exact dispersion relations for the EPOs are solved numerically and compared to the ones 
obtained by using approximate formulas for the electron susceptibility in the high- and low-frequency cases.
\end{abstract}

\maketitle

\section{Introduction}

The field of quantum plasma physics is becoming of increasing current
interest [\cite{Bonitz03,Manfredi05,Shukla06,Shaikh07,Crouseilles08,Serbeto08,Shukla09}], motivated by 
its potential applications in modern technology (e.g. metallic and semiconductor nanostructures-such 
as metallic nanoparticles, metal clusters, thin metal films, spintronics, nanotubes, quantum well and 
quantum dots, nano-plasmonic devices, quantum x-ray free-electron lasers, etc.). In dense quantum plasmas 
and in the Fermi gas of metals, the number densities of degenerate electrons are extremely high so 
that their wave functions overlap, and they therefore electrons obey the Fermi-Dirac statistics. 
The collective oscillations in quantum plasmas have been studied by several authors in the 
past [\cite{Klimontovich52,Bohm52,Bohm53,Pines1,Pines2,Ferrel}] with applications to the Fermi 
plasmas in metals and semiconductors, and to electrostatic oscillations in quantum pair plasmas [\cite{Mendonca08}].  
\cite{Watanabe} studied 
experimentally the Bohm-Pines dispersion relation of the electron plasma oscillations by measuring the energy 
loss of electrons by the excitation of collective modes in metals. The Fermi degenerate dense plasma may also 
arise when a pellet of hydrogen is compressed to many times the solid density in the fast ignition scenario for 
inertial confinement fusion [\cite{Azechi91,Azechi06,Son05,Lindl95,Tabak94,Tabak05}].  Since there is 
an impressive developments in the field of short pulse petawatt laser technology, it is highly likely 
that such plasma conditions can be achieved by intense laser pulse compression using  powerful x-ray pulses. 
Here ultrafast x-ray Thomson scattering techniques can be used to measure the features of laser enhanced 
plasma lines, which will, in turn, give invaluable informations regarding the equation of state of shock 
compressed dense matters.  Recently, spectrally resolved x-ray scattering
measurements [\cite{Kritcher08,Lee09}] have been performed in dense plasmas allowing accurate measurements 
of the electron velocity distribution function, temperature, ionization state, and of plasmons in the
warm dense matter regime [\cite{Glenzer07}]. This novel technique promises to access the degenerate, 
the closely coupled, and the ideal plasma regime, making it possible to investigate extremely dense 
states of matter, such as the inertial confinement fusion fuel during compression, reaching super-solid densities.

In this paper, we present a study of the dispersion properties of electrostatic oscillations in a dense quantum 
plasma, by employing the Wigner-Poisson model. We point out the differences between different regimes 
comprising the relatively low density regime of semiconductor plasmas, and the higher density regimes 
corresponding to metallic electron densities and laser compressed plasmas, as well as plasmas in 
dense astrophysical objects such as white dwarf stars.

\section{Derivation of the dispersion relation for the Wigner-Poisson system}

We here present a derivation of the dispersion relation for electrostatic waves in a
degenerate quantum plasma. The electron dynamics is governed by the Wigner equation

\begin{eqnarray}
  \nonumber
  &&\frac{\partial f_1}{\partial t}+{\bf v}\cdot \nabla f_1
  =-\frac{i e m_e^3}{(2\pi)^3\hbar^4}\int \int d^3\lambda d^3v'
  \exp\left[i \frac{m_e}{\hbar} ({\bf v}-{\bf v}')
  \cdot{\boldsymbol{\lambda}}\right]
  \\
  &&\times\left[
  \phi_1\left({\bf x}+\frac{\boldsymbol{\lambda}}{2},t\right)
 -\phi_1\left({\bf x}-\frac{\boldsymbol{\lambda}}{2},t\right)\right]f_0({\bf v}'),
  \label{eq1}
\end{eqnarray}
where the electrostatic potential $\phi$ is given by the Poisson equation
\begin{equation}
  \nabla^2\phi_1=\frac{e}{\epsilon_0} \bigg(\int f_1 d^3 v-n_{i1}\bigg).
  \label{eq2}
\end{equation}
Here $e$ is the magnitude of the electron charge, $m_e$ is the electron mass, $\hbar$ is the Planck 
constant divided by $2 \pi$, and $\epsilon_0$ is the permittivity of free space. Furthermore,
$f_0$ and $n_0$ denote the equilibrium electron distribution function and the electron number density, 
respectively, while $f_1$, $\phi_1$ and $n_{i1}$ denote the perturbed electron distribution function, 
the electrostatic potential, and the ion number density, respectively. 

Assuming that $f_1$, $\phi_1$ and $n_{i1}$ are proportional to $\exp(-i\omega t+i {\bf k} \cdot {\bf x})$, 
where $\omega$ is the frequency and ${\bf k}$ is the wave vector, we obtain from Eq. (\ref{eq1}) and (\ref{eq2}), 
respectively,

\begin{eqnarray}
  \nonumber
  &&(\omega-{\bf k}\cdot{\bf v})f_1
  =\frac{e m_e^3}{(2\pi)^3\hbar^4}\int \int d^3\lambda d^3v'
  \exp\left[i \frac{m_e}{\hbar} ({\bf v}-{\bf v}')
  \cdot{\boldsymbol{\lambda}}\right]
  \\
  &&\times \left[
    e^{i {\bf k}\cdot\boldsymbol{\lambda}/2}-e^{-i {\bf k}\cdot\boldsymbol{\lambda}/2}
  \right]f_0({\bf v}')\phi_1(\omega,{\bf k}),
  \label{eq_B3}
\end{eqnarray}
\begin{equation}
  k^2\phi_1=-\frac{e}{\epsilon_0}\bigg(\int f_1 d^3 v-n_{i1}\bigg).
  \label{eq_B4}
\end{equation}

Since ions are non-degenerate in quantum plasmas, we have for $\omega \gg k V_{Ti}$,

\begin{equation}
  n_{i1}=-\frac{\epsilon_0 k^2}{e}\chi_i \phi,
  \label{ni1}
\end{equation}
where

\begin{equation}
\chi_i=-\frac{\omega_{pi}^2}{\omega^2}
\label{chi_i}
\end{equation}
is the ion susceptibility, $V_{Ti}$ is the ion thermal speed, and $\omega_{pi}$ is the ion plasma frequency.

Rewriting (\ref{eq_B3}) as

\begin{eqnarray}
  \nonumber
  &&(\omega-{\bf k}\cdot {\bf v})f_1
  =\frac{i e m_e^3}{(2\pi)^3\hbar^4}\int \int d^3\lambda d^3v'
  \\
  \nonumber
  &&\times\left\{
    \exp\left[\frac{m_e}{\hbar} ({\bf v}-{\bf v}')
  \cdot{\boldsymbol{\lambda}}+i {\bf k}\cdot\boldsymbol{\lambda}/2\right]
  \right.
  \\
  &&\left.
  - \exp\left[i \frac{m_e}{\hbar} ({\bf v}-{\bf v}')
  \cdot{\boldsymbol{\lambda}}-i {\bf k}\cdot\boldsymbol{\lambda}/2\right]
  \right\}f_0({\bf v}')\phi_1(\omega,{\bf k}),
\end{eqnarray}
and performing the integration over $\boldsymbol{\lambda}$ space, we have

\begin{eqnarray}
  \nonumber
  &&(\omega-{\bf v}\cdot{\bf k})f_1
  =\frac{e m_e^3}{\hbar^4}\int d^3v'
   \left\{
    \delta\left[\frac{m_e}{\hbar} ({\bf v}-{\bf v}')+\frac{\bf k}{2}\right]
    \right.
    \\
    &&\left.
   -\delta\left[\frac{m_e}{\hbar} ({\bf v}-{\bf v}')-\frac{\bf k}{2}\right]
  \right\}f_0({\bf v}')\phi_1(\omega,{\bf k}),
\end{eqnarray}
where $\delta$ is the Dirac delta function. Now, the integration can be performed over ${\bf v}'$ space,
obtaining the result

\begin{equation}
  (\omega-{\bf k}\cdot{\bf v})f_1
  =\frac{e}{\hbar}
   \left[
     f_0\left({\bf v}+\frac{\hbar{\bf k}}{2m_e}\right)
    -f_0\left({\bf v}-\frac{\hbar{\bf k}}{2m_e}\right)
  \right]\phi_1(\omega,{\bf k}).
  \label{eq_B7}
\end{equation}
Eliminating $n_{i1}$ and $f_1$ in (\ref{eq_B4}) with the help of (\ref{ni1}) and (\ref{eq_B7}), we obtain the 
dispersion relation 

\begin{equation}
1+\chi_e+\chi_i=0,
\label{disp_e}
\end{equation}
where the ion susceptibility is given by (\ref{chi_i}) and the electron susceptibility is given by
\begin{equation}
  \chi_e=-\frac{4 \pi e^2k^2}{\hbar}
   \int \left[
     \frac{f_0\left({\bf v}+\frac{\hbar{\bf k}}{2m_e}\right)}{(-\omega+{\bf k}\cdot {\bf v})}
    -\frac{f_0\left({\bf v}-\frac{\hbar{\bf k}}{2m_e}\right)}{(-\omega+{\bf k}\cdot {\bf v})}
  \right] d^3 u.
  \label{chi_e1}
\end{equation}
Suitable changes of variables in the two terms in square brackets in Eq. (\ref{chi_e1}) now give

\begin{equation}
  \chi_e=-\frac{4 \pi e^2 k^2}{\hbar}
   \int \left[
     \frac{1}{[-\omega+{\bf k}\cdot({\bf u}-\frac{\hbar{\bf k}}{2 m_e})]}
    -\frac{1}{[-\omega+{\bf k}\cdot ({\bf u}+\frac{\hbar{\bf k}}{2 m_e})]}
  \right]f_0({\bf u}) d^3 u,
\end{equation}
which can be rewritten as

\begin{equation}
  \chi_e=-\frac{4 \pi e^2}{m_e}
   \int
     \frac{f_0({\bf u})}{(\omega-{\bf k}\cdot{\bf u})^2-\frac{\hbar^2 k^4}{4 m_e^2}} d^3 u.
  \label{eq_B10}
\end{equation}
This expression was also been derived by \cite{Bohm53} by using a series of canonical transformations 
of the Hamiltonian of the system [see for example the dispersion relation (57) in their paper], 
and by \cite{Ferrel} by using the method of self-consistent fields. 

We now choose a coordinate system such that the $x$ axis is aligned with the wave vector ${\bf k}$. 
Then, (\ref{eq_B10}) takes the form

\begin{equation}
  \chi_e=-\frac{4 \pi e^2}{m_e}
   \int
     \frac{f_0({\bf u})}{(\omega-k u_x)^2-\frac{\hbar^2 k^4}{4 m_e^2}} d^3 u.
  \label{eq_B11}
\end{equation}

We next consider a dense plasma with degenerate electrons in the zero temperature limit. 
Then, the background distribution function takes the simple form

\begin{equation}
  f_0=\left\{\begin{array}{cc}
  2\left(\frac{m_e}{2\pi \hbar}\right)^3, & |{\bf u}|\leq V_{Fe} \\
  0, & \mbox{elsewhere,}
  \end{array}
  \right.
  \label{flat}
\end{equation}
where $V_{Fe}=(2{\cal E}_{Fe}/m_e)^{1/2}$ is the speed of an electron on the Fermi surface, and
${\cal E}_{Fe}=(3\pi^2 n_0)^{2/3}\hbar^2/(2m_e)$ is the Fermi energy. The integration in (\ref{eq_B11}) 
can be performed over velocity space perpendicular to $u_x$, using cylindrical coordinate in $u_y$ and $u_z$, 
obtaining the result

\begin{equation}
  \chi_e=-\frac{4 \pi e^2}{m_e}
   \int
     \frac{F_0(u_x)}{(\omega-k u_x)^2-\frac{\hbar^2 k^4}{4 m_e^2}} d u_x,
  \label{eq_B13}
\end{equation}
where
\begin{eqnarray}
\nonumber
&&F_0(u_x)=\int\int f_0({\bf u})du_y du_z=2\pi\int_0^{\sqrt{V_{Fe}^2-u_x^2}} 2
\left(\frac{m_e}{2\pi \hbar}\right)^3u_\perp\,du_\perp
\\
&&=\left\{\begin{array}{cc}
2\pi\left(\frac{m_e}{2\pi \hbar}\right)^3(V_{Fe}^2-u_x^2), & |u_x|\leq V_{Fe} \\
0, & \mbox{elsewhere.}
\end{array}
\right.
\label{F0}
\end{eqnarray}

It is interesting to note that the distribution, which is flat-topped in three-dimensions becomes parabola-shaped 
in the remaining velocity dimension after the integration over the two perpendicular velocity dimensions. 
Hence, the electron distribution function $F_0(u_x)$ in (\ref{F0}) may support
Landau damping if the pole of the denominator in (\ref{eq_B13}) falls into the range of
negative slope of $F_0(u_x)$ in velocity space.
Equation (\ref{eq_B13}) can be written as

\begin{eqnarray}
  \nonumber
  &&\chi_e
  =-\frac{8 \pi^2 e^2}{m_e}\left(\frac{m_e}{2\pi \hbar}\right)^3
   \int_{-V_{Fe}}^{V_{Ve}}
     \frac{V_{Fe}^2-u_x^2}{(\omega-k u_x)^2-\frac{\hbar^2 k^4}{4 m_e^2}} d u_x
     \\
   &&=\frac{3\omega_{pe}^2}{4 V_{Fe}^3}
   \int_{-V_{Fe}}^{V_{Ve}}
     \frac{V_{Fe}^2-u_x^2}{(\omega-k u_x)^2-\frac{\hbar^2 k^4}{4 m_e^2}} d u_x.
  \label{eq_B15}
\end{eqnarray}

Performing the integration over velocity space, we have from (\ref{eq_B15})

\begin{eqnarray}
  \nonumber
  &&\chi_e=\frac{3\omega_{pe}^2}{4 k^2 V_{Fe}^2}\left\{
  2-\frac{m_e}{\hbar k V_{Fe}}\left[V_{Fe}^2-\left(\frac{\omega}{k}+\frac{\hbar k}{2 m_e}\right)^2\right]
  \log\left|
  \frac{
  \frac{\omega}{k}-V_{Fe}+\frac{\hbar k}{2 m_e}
  }{
  \frac{\omega}{k}+V_{Fe}+\frac{\hbar k}{2 m_e}
  }
  \right|
  \right.
  \\
  &&\left.
  +\frac{m_e}{\hbar k V_{Fe}}\left[V_{Fe}^2-\left(\frac{\omega}{k}-\frac{\hbar k}{2 m_e}\right)^2\right]
  \log\left|
  \frac{
  \frac{\omega}{k}-V_{Fe}-\frac{\hbar k}{2 m_e}
  }{
  \frac{\omega}{k}+V_{Fe}-\frac{\hbar k}{2 m_e}
  }
  \right|
  \right\}.
  \label{eq_B18}
\end{eqnarray}
In the derivation of (\ref{eq_B18}), we have assumed that the waves are only weakly damped, so that, 
when integrating over poles, only the principal parts of the integrals are kept.
In the limit $\hbar k /m_e\rightarrow 0$, we have from (\ref{eq_B18})

\begin{equation}
  \chi_e=\frac{3\omega_{pe}^2}{k^2 V_{Fe}^2}\left(
  1-\frac{\omega}{2k V_{Fe}}\log\left|\frac{\omega+k V_{Fe}}{\omega-k V_{Fe}}\right|
  \right),
  \label{eq_B16}
\end{equation}
where it holds that $\omega$ is real and $\omega/k>V_{Fe}$. Hence in this
"semi-classical" limit, we do not have Landau damping. [We call (\ref{eq_B16}) "semi-classical", since 
it can be derived from the Vlasov equation for electrons using the flat-topped background electron distribution function given by (\ref{flat}).]

\section{Electron oscillations}
We here consider high-frequency ($\omega\gg\omega_{pi}$) oscillations so that $\chi_i\ll 1$ in (\ref{chi_i}).
Then, expanding (\ref{eq_B18}) for small wavenumbers up to terms containing $k^4$, we obtain
from (\ref{disp_e}) the dispersion relation

\begin{equation}
  \omega^2=\omega_{pe}^2+\frac{3}{5}k^2 V_{Fe}^2+(1+\alpha)\frac{\hbar^2 k^4}{4 m_e^2},
  \label{eq_B19}
\end{equation}
where $\alpha=(48/175)m_e^2 V_{Fe}^4/\hbar^2\omega_{pe}^2\approx 2.000 (n_0 a_0^3)^{1/3}$ and 
$a_0=\hbar^2/m_e e^2\approx 53\times 10^{-10}\,\mathrm{cm}$ is the Bohr radius. For a typical metal 
such as gold, which has a free electron number density of $n_0=5.9\times10^{22}\,\mathrm{cm}^{-3}$, 
we would have $\alpha\approx 0.4$. For the free electron density in semiconductors, which is many orders 
of magnitude less than in metals, $\alpha$ is much smaller and can safely be dropped compared to unity. 
However, for electron plasma oscillations in dense matters, $\alpha$ could be larger than unity. 
It should be noted that the term proportional to $\alpha$ in (\ref{eq_B19}) was not discussed 
by \cite{Bohm53} and others, but was, however, obtained and discussed by \cite{Ferrel} in his study 
of collective electron oscillations in metals [see Eq. (10) in his paper, where, in his notation, it should be
$(\Delta v^2/v_0^2)^2=12/175$].

It is interesting to note that (\ref{eq_B18}) \emph{may admit} Landau damping above a certain critical 
wavenumber, $k>k_{cr}$ and corresponding frequency $\omega>\omega_{cr}$. This occurs if the denominator 
in the integral of (\ref{eq_B15}) vanishes within the integration limits $u_x=\pm V_{Fe}$. 
For the critical wavenumber and frequency, we have

\begin{equation}
\omega=\omega_{cr}=k_{cr} V_{Fe}+\frac{\hbar k_{cr}^2}{2} m_e.
\label{critical1}
\end{equation}
Inserting this expression into (\ref{eq_B18}) we note that the term involving the logarithm on 
the second line of (\ref{eq_B18}) vanishes, and we obtain the critical wavenumber $k=k_{cr}$ from

\begin{equation}
  1+\frac{3\omega_{pe}^2}{4 k_{cr}^2 V_{Fe}^2}\bigg[
  2-\bigg(2+\frac{\hbar k_{cr}}{m_e V_{Fe}}\bigg)
  \log
  \bigg(
    1+\frac{2 m_e V_{Fe}}{\hbar k_{cr}}
  \bigg)
  \bigg]=0.
  \label{critical2}
\end{equation}
A careful examination of the dispersion relation for $k>k_{cr}$, should involve Landau contours to correctly 
take into account Landau damping. Here we are interested in low- and high-frequency waves in the weakly 
damped regime and have postponed the investigation of Landau damping of the system to future studies.

\section{Ion oscillations}

In a quantum plasma system composed of mobile ions and inertialess electrons, we have possibility 
of low-phase speed (in comparison with the Fermi electron thermal speed) ion-acoustic-like oscillations.
For low-frequency ($\omega\ll k V_{Fe}$) waves, we have from (\ref{eq_B18})

\begin{equation}
  \chi_e=\frac{3\omega_{pe}^2}{2 k^2 V_{Fe}^2}
  \bigg[
    1-\frac{m_e}{\hbar k V_{Ve}}\bigg( V_{Fe}^2-\frac{\hbar^2 k^2}{4 m_e^2}\bigg)\log
    \left|
    \frac{V_{Fe}-\frac{\hbar k}{2m_e}}{V_{Fe}+\frac{\hbar k}{2 m_e}}
    \right|
  \bigg].
  \label{chi_e}
\end{equation}
For small wavenumbers $\hbar k\ll m_e V_{Fe}$,  we have the approximate
electron susceptibility, up to terms containing factors of $k^4$,

\begin{equation}
  \chi_e\approx \frac{3\omega_{pe}^2}{k^2 V_{Fe}^2+\hbar k^4/12 m_e^2}.
  \label{chi_e2}
\end{equation}

Using the dispersion relation
\begin{equation}
  {\varepsilon}(\omega,k)=1+\chi_e+\chi_i=0,
\end{equation}
we employ (\ref{chi_i}) to obtain the frequency of ion acoustic waves as

\begin{equation}
  \omega=\frac{\omega_{pi}}{(1+\chi_e)^{1/2}},
  \label{omega_ion}
\end{equation}
with $\chi_e$ given by (\ref{chi_e}) or (\ref{chi_e2}). For $\chi_e$ given by the approximate
expression (\ref{chi_e2}), we have

\begin{equation}
  \omega=\frac{\omega_{pi}(k^2 V_{Fe}^2+\hbar^2 k^4/12 m_e^2)^{1/2}}
{(3\omega_{pe}^2+k^2V_{Fe}^2+\hbar^2 k^4/12 m_e^2)^{1/2}}
=\frac{k C_s (1+\hbar^2 k^2/12 m_e^2 V_{Fe}^2)^{1/2}}
{(1+k^2 V_{Fe}^2/3\omega_{pe}^2+\hbar^2 k^4/36 m_e^2 \omega_{pe}^2)^{1/2}},
  \label{ion_acoustic}
\end{equation}
where $C_s=\sqrt{m_e V_{Fe}^2/ 3 m_i}$ is the Fermi ion acoustic speed. 
We note that $\omega\rightarrow \omega_{pi}$ as $k\rightarrow\infty$.

\section{Density regimes of the system}

We note that there is a critical density parameter in the system. When the inter-particle distance is smaller than 
the Bohr radius, then the quantum statistical pressure dominates the wave dynamics, while in the opposite case, 
the quantum tunneling effects become important when the wavelength is comparable to the inter-particle distance. 
This can be seen by normalize the system such that $\omega/\omega_{pe}=\Omega$ and $k V_{Fe}/\omega_{pe}=K$. 
Then, Eq. (\ref{eq_B18}) takes the form

\begin{eqnarray}
  \nonumber
  &&\chi_e=\frac{3}{4 K^2}\left\{
  2-\frac{\beta}{K}\left[1-\left(\frac{\Omega}{K}+\frac{K}{2 \beta}\right)^2\right]
  \log\left|
  \frac{
  \frac{\Omega}{K}-1+\frac{K}{2\beta}
  }{
  \frac{\Omega}{K}+1+\frac{K}{2\beta}
  }
  \right|
  \right.
  \\
  &&\left.
  +\frac{\beta}{K}\left[1-\left(\frac{\Omega}{K}-\frac{K}{2\beta}\right)^2\right]
  \log\left|
  \frac{
  \frac{\Omega}{K}-1-\frac{K}{2\beta}
  }{
  \frac{\Omega}{K}+1-\frac{K}{2\beta}
  }
  \right|
  \right\},
  \label{eq_B20}
\end{eqnarray}
where

\begin{equation}
  \beta=\frac{m_e V_{Fe}^2}{\hbar\omega_{pe}}=3^{2/3}\pi^{5/6}(a_0^3 n_0)^{1/6}.
\end{equation}
In the scaled variables, (\ref{eq_B16}) and (\ref{eq_B19}) take the form

\begin{equation}
  \chi_e=\frac{3}{K^2}\left(
  1-\frac{\Omega}{2K}\log\left|\frac{\Omega+K}{\Omega-K}\right|
  \right),
  \label{eq_semi_scaled}
\end{equation}
and

\begin{equation}
  \Omega^2=1+\frac{3}{5}K^2+\bigg(\frac{1}{\beta^2}+\frac{48}{175}\bigg)\frac{ K^4}{4},
\end{equation}
respectively. We note that the limit $\hbar k/m_e\rightarrow 0$ to obtain (\ref{eq_B16}) from (\ref{eq_B18}) 
corresponds to $\beta\rightarrow \infty$ to obtain (\ref{eq_semi_scaled}) from (\ref{eq_B20}).
For $\beta\gg 1$, the quantum statistical pressure dominates, while for $\beta\ll 1$, the quantum 
tunneling effects dominate.
Considering the value of $\beta$ for different physical systems, we note that $\beta=0.1$ corresponds to 
relatively low density degenerate plasma such as in semiconductors, while $\beta=1$ corresponds to typical 
free electron densities in metals. The high density case $\beta=10$ corresponds to high density matter 
which may be obtained in laser compression schemes or which exist in white dwarf stars. Even though 
we have formally considered the semiclassical limit $\beta\rightarrow\infty$, it should be kept in mind 
that an upper limit for the validity of our theory is when the electron density becomes high enough that 
the Fermi speed $V_{Fe}$ becomes comparable with the speed of light. In this limit, the inter-particle distance $n^{-1/3}$ 
approaches the Compton length $\lambda_C=2\pi\hbar/m_e c\approx 2.4\times 10^{-12}$, and we have an 
electron number density of the order $10^{35}\,\mathrm{m}^{-3}$, corresponding to $\beta\approx 27$. 
For larger values of $\beta$, the equilibrium equation of state for the electrons [\cite{Chandra}] change 
from $P=({2}/{5}){\cal E}_F n_0 ({n_e}/{n_0})^{5/3}$ to $P=({3}/{\pi})^{1/3}({4\pi c h}/{8})n_e^{4/3}$.
For this case, we need to include relativistic effects in the electron susceptibility.

In the normalized variables, the condition (\ref{critical2}) for the critical wavenumber for the
limit between undamped and Landau damped high-frequency waves is given by

\begin{equation}
  1+\frac{3}{4 K_{cr}^2}\bigg[
  2-\bigg(2+\frac{K_{cr}}{\beta}\bigg)
  \log
  \bigg(
    1+\frac{2\beta}{K_{cr}}
  \bigg)
  \bigg]=0,
\end{equation}
and the normalized critical frequency is obtained from (\ref{critical1}) as

\begin{equation}
  \Omega_{cr}=K_{cr}+K_{cr}^2/2\beta,
\end{equation}
where $K_{cr}=k_{cr} V_{Fe}/\omega_{pe}$ and $\Omega_{cr}=\omega_{cr}/\omega_{pe}$.

Finally, for the low-frequency case, the electron susceptibilities (\ref{chi_e})
is in the normalized variables given by
\begin{equation}
  \chi_e=\frac{3}{2 K^2}
  \bigg[
    1-\frac{\beta}{K}\bigg( 1-\frac{K^2}{4\beta^2}\bigg)\log
    \left|
    \frac{1-\frac{K}{2\beta}}{1+\frac{K}{2\beta}}
    \right|
  \bigg].
\end{equation}
and the expression for small wavenumbers (\ref{chi_e2}) is given by
\begin{equation}
  \chi_e\approx \frac{3}{K^2+\hbar K^4/12\beta^2}.
\end{equation}

\section{Numerical results}

\begin{figure}
  \includegraphics[width=10cm]{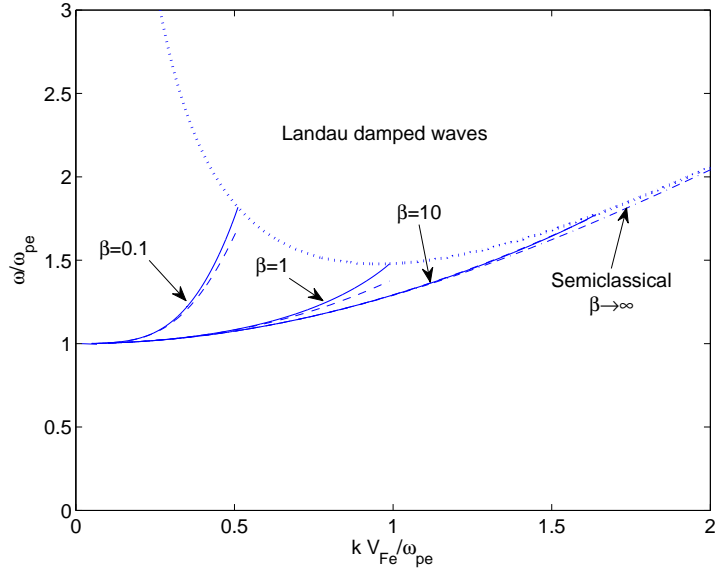}
  \caption{Dispersion curves ($\omega$ versus $k$) for different values of $\beta=m_e V_{Fe}^2/\hbar \omega_{pe}$. 
The solid curves show solutions of (\ref{disp_e}) using the exact susceptibility (\ref{eq_B18}), the dashed curves 
show the expanded solution (\ref{eq_B19}), and the dash-dotted curve shows solutions of (\ref{disp_e}) using 
the "semi-classical" electron susceptibility (\ref{eq_B16}). The dotted curve indicates the border between 
undamped waves and Landau damped waves, given by (3.2) and (3.3) [or (5.5) and (5.6)].}
\end{figure}

\begin{figure}
  \includegraphics[width=10cm]{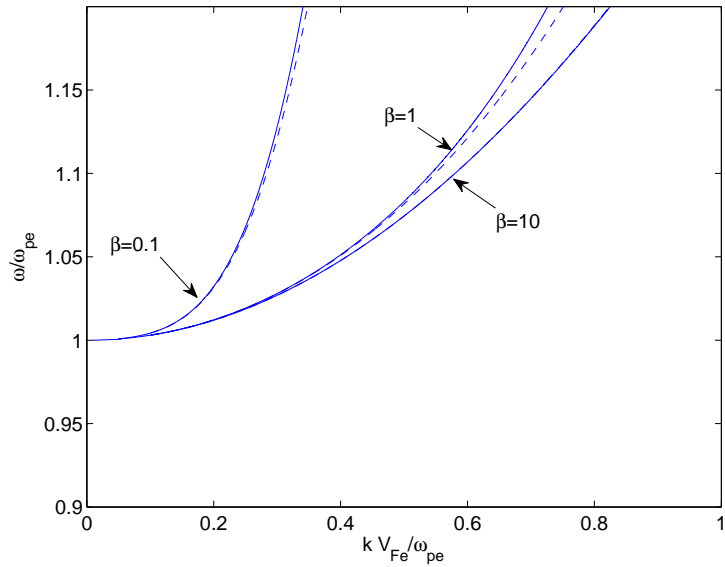}
  \caption{Closeup of the dispersion curves in Fig. 1.}
\end{figure}

\begin{figure}
  \includegraphics[width=10cm]{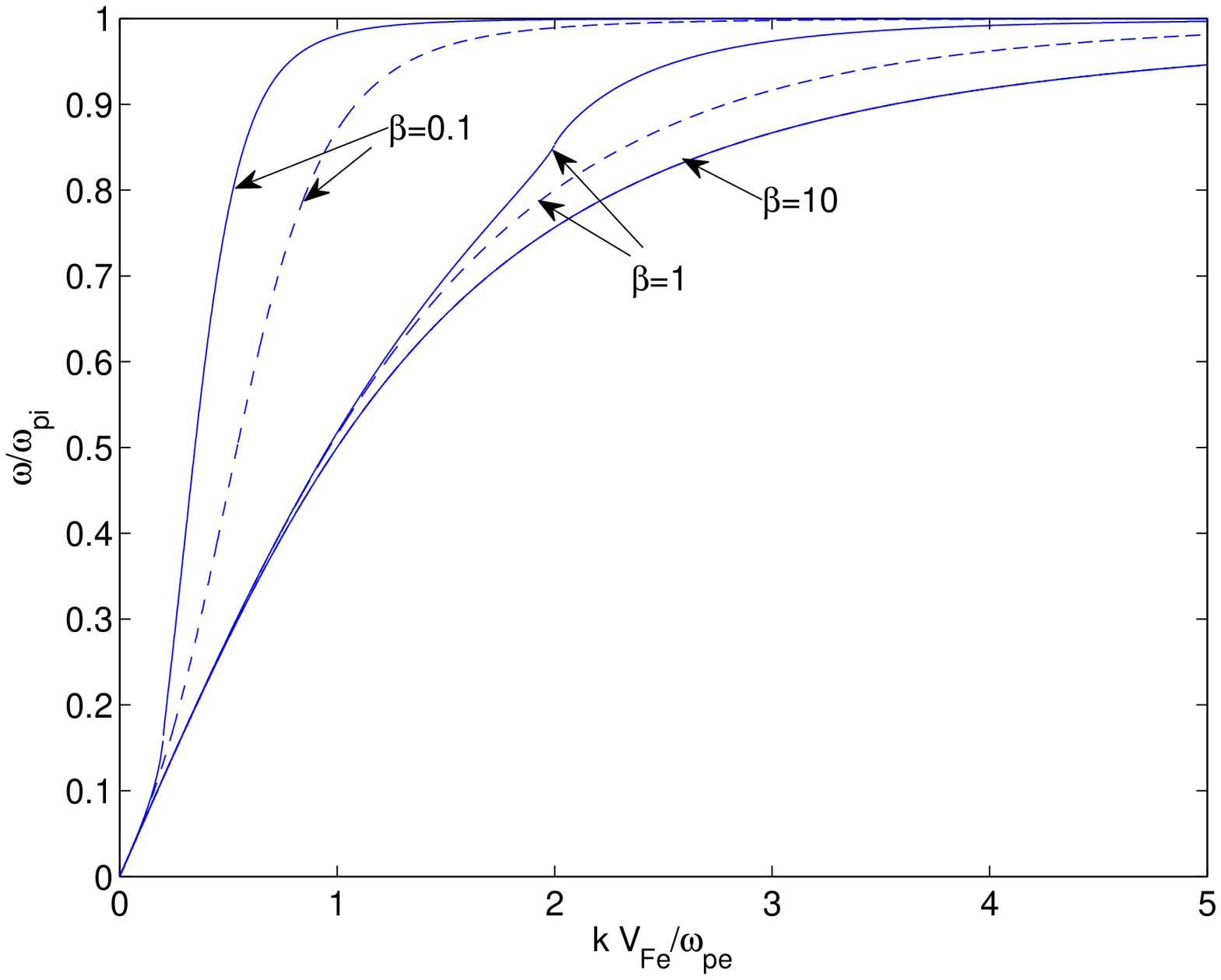}
  \caption{Dispersion curves ($\omega$ versus $k$) for the low-frequency ion oscillations, for different 
values of $\beta=m_e V_{Fe}^2/\hbar \omega_{pe}$. The solid curves show the wave frequency given 
by (\ref{omega_ion}) using the low-frequency electron susceptibility (\ref{chi_e}), while the dashed 
curves uses the approximate electron susceptibility (\ref{chi_e2}). For $\beta=10$ the solid and dashed curves
are indistinguishable.}
\end{figure}

In Figs. 1 and 2, we show dispersion curves for the high-frequency ($\omega\gg\omega_{pi}$) waves for 
different values of $\beta$, obtained from the solutions of the dispersion relation (\ref{disp_e}), 
by using the electron susceptibility (\ref{eq_B18}), as well as the expansion (\ref{eq_B19}) and the 
limiting semi-classical case (\ref{eq_B16}). We have also indicated the border between undamped and 
Landau damped waves, obtained from (\ref{critical1}) and (\ref{critical2}). 
We note that the dispersion curve 
for the semiclassical case in Fig 1 always lies in the undamped regime, 
below the border between undamped and Landau damped waves. 
For the undamped waves, the expansion (\ref{eq_B19}) approximates the exact dispersion relation within 
a few percent, and can therefore be used instead of (\ref{eq_B18}) for most cases. This holds especially 
for small wavenumbers, as can be seen in the closeup in Fig. 2.

The dispersion curves for the low-frequency ion-acoustic oscillations are plotted in Fig. 3, were we 
have depicted the $\omega$ in (\ref{omega_ion}) for $\chi_e$ given by the exact expression (\ref{chi_e}) and
the approximate expansion (\ref{chi_e2}), and for different values of $\beta$.
We note that the dispersion curves show agreement at small
wavenumbers but deviate significantly for larger wavenumbers in the case $\beta=0.1$, while the
agreement is better for $\beta=1$ and excellent for $\beta=10$. It should be kept in mind that
when $\hbar k \sim m_e V_{Fe}$ then the wavelength of the oscillations is comparable to the inter-particle distance, 
and there will be corrections due to the discrete nature of the ion background. Thus, the theory is not valid 
for oscillations with $\hbar k \gg m_e V_{Fe}$.

\section{Conclusions}

In this paper, we have here studied the dispersion properties of electrostatic oscillations in quantum 
plasmas for different parameters ranging from semiconductor plasmas, to typical metallic electron densities 
and to densities corresponding to compressed matter and dense astrophysical objects. We have derived a 
simplified expansion that accurately approximates the exact dispersion relation for small wavenumbers. 
The possibility of Landau damping due to quantum tunneling effects at large wavenumbers has also 
been discussed, and conditions for Landau damping has been derived. The present results should be useful
in understanding the salient features of electrostatic plasma oscillations in dense plasmas with degenerate
electrons. The latter are encountered in metals, in highly compressed intense laser-solid density 
plasma experiments, and in compact astrophysical objects (e.g. interior of white dwarf stars).

\begin{acknowledgments}
This work was supported by the Deutsche Forschungsgemeinschaft through the project SH21/3-1 of the 
Research Unit 1048, and by the Swedish Research Council (VR).
\end{acknowledgments}

\end{document}